# Systematic Absences of Optical Phonon Modes in Phonon Dispersion Measured by Electron Microscopy

Aowen Li (李傲雯)[1], Paul Zeiger[2], Zuxian He (何祖贤)[2], Mingquan Xu (许名权)[1], Stephen J. Pennycook[1], Ján Rusz[2,*], Wu Zhou (周武)[1,*]

[1]School of Physical Sciences and CAS Key Laboratory of Vacuum Physics, University of Chinese Academy of Sciences, Beijing 100049, P. R. China

[2]Department of Physics and Astronomy, Uppsala University, Box 516, Uppsala 75120, Sweden

*Corresponding author. Email: wuzhou@ucas.ac.cn (W.Z.); jan.rusz@physics.uu.se (J.R.)

**Abstract**

Phonon dispersion is widely used to elucidate the vibrational properties of materials. As an emerging technique, momentum-resolved vibrational spectroscopy in scanning transmission electron microscopy (STEM) offers an unparalleled approach to explore *q*-dependent phonon behavior at local structures. In this study, we systematically investigate the phonon dispersion of monolayer graphene across several Brillouin zones (BZs) using momentum-resolved vibrational spectroscopy and find that the optical phonon signals vanish at the Γ points with indices (*hk*0) satisfying $h + 2k = 3n$ (*n* denoted integers). Theoretical analysis reveals that the observed phenomena arise from the complete destructive interference of the scattered waves from different basis atoms. This observation, corroborated by the study of diamond, should be a general characteristic of materials composed of symmetrically equivalent pairs of the same elements. Moreover, our results emphasize the importance of multiple scattering in interpreting the vibrational signals in bulk materials. We demonstrate that the systematic absences and dynamic effects, which have not been much appreciated before, offer new insights into the experimental assessment of local vibrational properties of materials.

Phonon dispersion is crucial for describing the phonon-mediated behavior of materials. However, measuring phonon dispersions at local structures remains challenging in most vibrational spectroscopy methods, such as optical spectroscopy [1-5], inelastic X-ray or neutron scattering (IXS or INS) [6-8] and high resolution electron energy-loss spectroscopy (HREELS) [9,10], which typically suffer from insufficient spatial or momentum resolution. For example, HREELS has been utilized to measure the surface phonon dispersion of epitaxially grown monolayer graphite, albeit with limited spatial resolution [11]. Recently, significant advances in monochromated scanning transmission electron microscopy (STEM) have provided a unique approach to directly assess the local vibrational properties of lattice imperfection and their correlation with the atomic structure [12-27]. Vibrational spectroscopy in STEM could balance the momentum, spatial, and energy resolution, enabling the detection of phonon modes and dispersion at structural imperfections [15,18,22,28-30]. It is worth noting that phonon dispersion obtained through momentum-resolved vibrational spectroscopy in STEM exhibits variable intensities along phonon branches and demonstrates different features on the same high-symmetry points in different Brillouin zones (BZs) [31]. The interpretation of such features should take into consideration the cross-section of inelastic electron scattering [28,29,32], which is not included in any calculations of only the phonon dispersion.

To probe the material properties using vibrational spectroscopy in STEM, it is necessary to establish a comprehensive understanding of the interaction between the incident electrons and crystal lattices. Several theoretical methods have been developed for this purpose [29,32-39]. In particular, the method derived from the van Hove scattering formalism, originally designed for INS [40], has gained popularity for elucidating vibrational signals emerging at local structures [13,16,18,19]. The frequency-resolved frozen phonon multi-slice (FRFPMS) method, leveraging molecular dynamics and extending the frozen phonon approximation, offers inherently high computational efficiency and accommodates the effects of dynamical diffraction to interpret experimental vibrational intensities [30,31,41,42]. Complementary with theoretical simulation, Senga *et al.* discovered that vibrational signals in graphite and graphene vanish in the long-wavelength limit ($q \rightarrow 0$) because of the perfect screening of the ionic charge by the valence density in a semimetal [29]. However, since the perfect screening does not hold in a scattering process with a finite momentum transfer, the behavior of vibrational signals at the higher order Γ points is anticipated to differ from that observed at the central Γ point, thus deserving a comprehensive investigation.

Here, we employed momentum-resolved vibrational spectroscopy in STEM to investigate the vibrational signals within different BZs in monolayer graphene and diamond. The experiments were performed with a convergence semi-angle of 3.5 mrad under 60 kV, resulting in a probe size of ~1.2 nm and a diffraction spot radius of 0.45 Å$^{-1}$. Due to the extremely weak vibrational signals of single-layered graphene, especially within the momentum space beyond the first BZ, a slot-type EELS aperture with a collection range of 7×112 mrad$^2$ was used to facilitate high-efficiency parallel acquisitions along the Γ-K-M-K-Γ direction in momentum space [43]. It is also worth mentioning that a direct electron EELS detector was employed to eliminate readout noise in our data [44]. The energy resolution in the experiments is around 16 meV, sufficient to distinguish phonon modes in the acquired phonon dispersions. From these experimental conditions, we found that the optical phonon signals of graphene appear or disappear at different Γ points, forming a systematic pattern in momentum space. The consideration of cross-section of the inelastic scattering process within the van Hove formalism unveils that the complete destructive interference of electron waves inelastically scattered by different basis atoms in the graphene unit cell results in the disappearance of the optical phonon signals at specific Γ points, indicating that this phenomenon should be a general characteristic of materials composed of symmetrically equivalent pair(s) of identical elements. This hypothesis is confirmed by our joint experimental and theoretical investigation to diamond, another phase of elemental carbon. The results for diamond also suggest that multiple scattering can further modulate the vibrational signals of bulk materials. Our study indicates that the systematic absences of optical phonon signals and dynamical effects are critical factors in the interpretation of vibrational spectroscopy, providing new insights into the study of vibrational properties of materials.

Figures 1A-B present two distinct experimental phonon dispersion diagrams of graphene (Fig. S1A), collected at different regions in momentum space as indicated in Fig. 1C. Notably, despite both being collected along the Γ-K-M-K-Γ direction in momentum space and correlated to the same vibrational modes in theory, these two diagrams exhibit different features. For example, in the upper panel of Fig. S1B, which displays the vibrational spectra extracted from the corresponding M points of these two phonon dispersion diagrams, the red curve shows a prominent peak corresponding to the transverse acoustic (TA) phonon mode, whereas the blue curve exhibits higher intensity within the region of longitudinal acoustic (LA), longitudinal optical (LO) and transverse optical (TO) modes. The disparities in spectral characteristics observed at the M and K points in graphene align with the calculated vibrational spectra of hexagonal boron-nitride (h-BN), as reported by two of the coauthors of this work [31], which can be explained by considering the scalar product of the phonon polarization vector of the vibrational modes (***e***) and the momentum transfers of the inelastically scattered electrons (***q***) using the scattering cross-section within the van Hove formalism [29,32,40].

Regarding the Γ points (Fig. 1D), the red curve presents a prominent optical phonon peak at around 190 meV, while the blue curve lacks this feature. However, the variability in the visibility of vibrational signals at different Γ points cannot be attributed to the scalar product of ***e*** · ***q*** alone, because there is always at least one optical phonon branch that yields a nonzero scalar product ***e*** · ***q***, except at the central Γ point where ***q*** is zero. To explore the underlying mechanism governing the behavior of the optical phonon signals at the Γ points, we gathered a series of phonon dispersion diagrams collected from various momentum regions and checked the visibility of the signals in each diagram, as shown in Fig. S2. Based on the experimental findings and the six-fold structural symmetry inherent to graphene, a schematic in Fig. 1C denotes the Γ points in red or blue when the optical phonon peaks are visible or invisible, respectively, delineating a systematic pattern in momentum space.

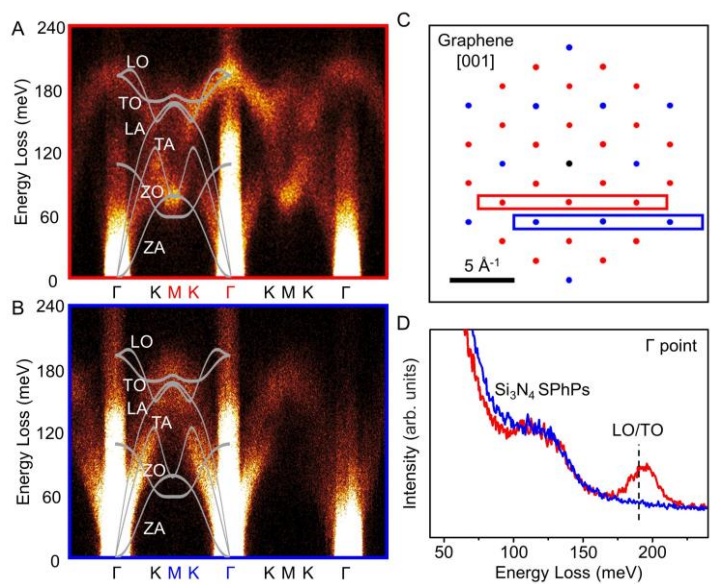

**Figure 1 Momentum-resolved vibrational spectroscopy of graphene at different momentum space regions.** (A, B) Experimental phonon dispersion diagrams of graphene, collected from the momentum spaces highlighted by red and blue rectangles in (C), respectively. (C) Schematic of graphene diffraction pattern along the [001] zone axis. The Γ points exhibiting or not exhibiting optical phonon mode signals are colored in red and blue, respectively. The central Γ point is shown in black. (D) Vibrational spectra extracted from the highlighted Γ points in (A-B). The spectra extracted from (A) and (B) are colored in red and blue, respectively. The dashed line in this figure indicates the energy of the TO/LO phonon modes at the Γ points. The signals in the energy-loss range of 100-150 meV in the spectra of the Γ points are attributed to the surface phonon polaritons (SPhPs) of the $Si_3N_4$ substrate, labeled as $Si_3N_4$ SPhPs.

To understand our experimental observations, we conducted simulations of the vibrational spectra of graphene using the FRFPMS method with parameters identical to those in our experiments (see Supplementary Materials for details) [31,41]. Figure 2 illustrates the chosen Γ points alongside their corresponding simulated vibrational spectra. In line with the notation employed in the experimental results, the simulated spectra from the red Γ points exhibited very strong optical phonon peaks, while the optical phonon signals in spectra from the blue Γ points almost vanish. The excellent agreement in the visibility of optical phonon signals at the Γ points between the simulated and experimental results is intriguing. Our results reveal that in monolayer graphene, the indices of Γ points ($hk0$), where optical phonon peaks are invisible, should satisfy:

$$h + 2k = 3n \tag{1}$$

where $h$, $k$, and $n$ are integers.

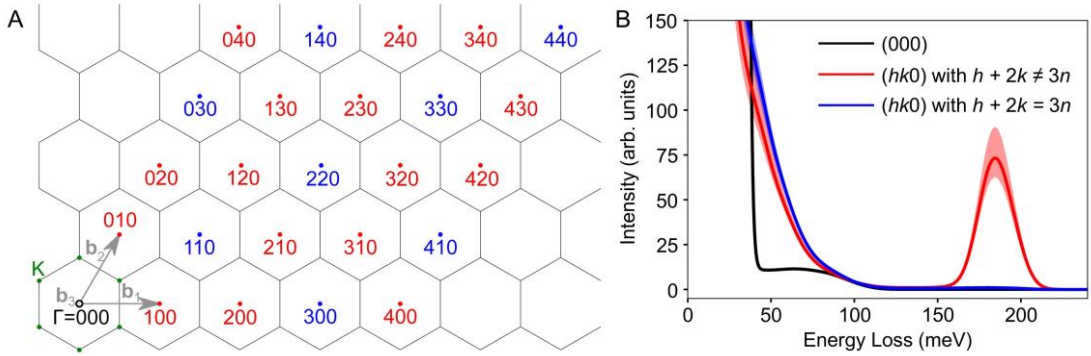

**Figure 2 Simulation of the graphene vibrational signals at different Γ points.** (A) Reciprocal lattice of graphene along the [001] zone axis. The selected Γ points in the simulations are indexed. (B) Simulated vibrational spectra at the Γ points highlighted in (A). The black solid line visualizes the spectrum at (000), while the red and blue solid lines correspond to the mean spectra of their corresponding groups. The shaded region around the mean indicates the spread of the corresponding group as it traces the minimum and maximum intensities.

To elucidate the factors contributing to the systematic absences of the graphene optical phonon signals at different Γ points, we considered the double differential scattering cross-section within the van Hove formalism [29,32,40]:

$$\frac{d^2\sigma}{d\Omega(\boldsymbol{q})dE} \propto \sum_\nu \frac{1+\langle n(\boldsymbol{q}_0,\nu)\rangle_T}{\omega(\boldsymbol{q}_0,\nu)} \delta\big(\omega - \omega(\boldsymbol{q}_0,\nu)\big) \left|\sum_\kappa e^{-i\boldsymbol{G}\cdot\boldsymbol{R}_\kappa} e^{-W_\kappa(\boldsymbol{q})} f_\kappa^e(\boldsymbol{q}) \frac{\boldsymbol{q}\cdot\boldsymbol{e}_\kappa(\boldsymbol{q}_0,\nu)}{\sqrt{M_\kappa}}\right|^2 \tag{2}$$

where $\nu$ is the phonon band index, $\boldsymbol{q} = \boldsymbol{G} + \boldsymbol{q}_0$ is the momentum transfer of the incident electron, $G$ is a reciprocal lattice vector and $q_0$ is a vector in the first Brillouin zone, $R_\kappa$ is the position of the $\kappa^{th}$ atom in the unit cell, $e^{-W_\kappa(q)}f_\kappa^e(q)$ is its associated thermally smeared scattering factor, and $M_\kappa$ is its mass. Furthermore $e_\kappa(q_0, \nu)$ is the phonon polarization vector at the $\kappa^{th}$ atom in mode $(q_0, \nu)$. It is notable that the two carbon (C) atoms in the unit cell of graphene have opposite phonon polarization vectors of the LO and TO modes at the Γ point ($e_2 = -e_1$), but identical thermally smeared scattering factor and the atomic mass. Thus, for optical phonon modes of graphene at Γ points, Eq. (2) can be expressed as follows:

$$\frac{d^2\sigma}{d\Omega(q)dE} \propto |q \cdot e_1(q_0, \nu)|^2 \left|e^{-iG \cdot R_1} - e^{-iG \cdot R_2}\right|^2 \tag{3}$$

For both LO and TO branches, the first factor of Eq. (3) only vanishes at $G = 0$, i.e., the central Γ point. At any other Γ point with $G \neq 0$, at least one of the eigenvectors of these branches $\nu$ has a nonzero component along the momentum transfer $q$, as we mentioned above.

The second factor of Eq. (3) has a similar form to the structure factor of graphene ($F_G \propto e^{iG \cdot R_1} + e^{iG \cdot R_2}$) but with a minus sign between the waves scattered at the different basis atoms. By considering the indices of Γ points ($hk0$) and the positions of the two basis atoms of graphene, the second factor of Eq. (3) can be rewritten as follows:

$$\left|e^{-i(hb_1+kb_2)\cdot R_1} - e^{-i(hb_1+kb_2)\cdot R_2}\right|^2 = 2\left[1 - \cos\left(\frac{2\pi(h+2k)}{3}\right)\right] \tag{4}$$

which equals zero when $h$ and $k$ satisfy $h + 2k = 3n$, where $n$ is an integer (see Supplementary Materials for derivation details). This consideration of the double differential scattering cross-section demonstrates that complete destructive interference of excitations of optical phonons from different atomic sites takes place at the Γ points with indices satisfying $h + 2k = 3n$, thereby giving rise to the systematic absences observed in our experiments. This highlights an importance of inter-atomic cross-terms in the transition potential formulation of the inelastic phonon scattering, where such cross-terms have been considered negligible [37].

Based on the above derivation, we conclude that the symmetrically equivalent pair(s)

of the same elements contributes significantly to the systematic absences in the optical phonon signals of graphene. However, for materials with dissimilar element pairs, their basis atoms have distinct weightings due to the variations in the magnitudes of their phonon polarization vectors, atomic masses and thermally smeared scattering factor. Therefore, in such cases, instead of complete disappearance, a reduction in the intensity of the vibrational signals should be expected, as shown in the simulated phonon dispersion of hexagonal boron nitride (h-BN) [31]. It should be mentioned that the discussion here is restricted to impact scattering and excludes any consideration of the SPhPs appearing at Γ points in experiment (Fig. S3) [45].

From the discussion above, it seems reasonable to infer that the systematic absences could also manifest in other materials that are composed of symmetrically equivalent pair(s) of the same elements, such as diamond. Diamond has a face-centered cubic (FCC) lattice, comprising of 8 C atoms in its conventional cell with a two-atom basis. The phonon eigenvectors at the Γ points for the two C atoms in the basis are antiparallel. Applying the same derivation as we have employed for graphene, the systematic absences of optical phonon peaks in diamond are determined by the rule that $h + k + l = 4n$ (see Supplementary Materials for derivation details).

To verify our theory, we collected phonon dispersions of diamond on two different zone axes of [001] and [110] with sample thicknesses of 56 and 43 nm, respectively (Fig. S4). Along the [001] axis, the optical phonon signals at around 165 meV disappear at the ($\bar{2}20$) and (220) spots but appear at the (020) spot, consistent with our theoretical derivation (Figs. 3A and C). However, along the [110] axis, spectra collected at all Γ points present optical phonon signals, including the ($\bar{2}20$) point that satisfies the $h + k + l = 4n$ rule (Figs. 3B and D). Furthermore, we observed that the kinematically forbidden spot (002) is visible in our experimental diffraction pattern along the [110] axis, which was captured from the same sample region as for the phonon dispersion measurement, suggesting strong multiple elastic scattering in the thick diamond sample. Multiple scattering could explain the discrepancies between the experimental results and theoretical derivation of the visibility of the optical phonon peaks, as this factor is not included in the theory originating from the van Hove formalism [40]. For example, in the [110] orientation, the optical phonon signals at the $(\bar{2}20)$ spot, which by the selection rule should vanish in diamond, can result from a combination of an inelastic scattering event to the $(\bar{1}11)$ spot and a subsequent elastic Bragg scattering event to the $(\bar{1}1\bar{1})$ spot. In contrast, introducing an optical phonon peak for the $(\bar{2}20)$ spot in the [001] orientation from such a combination of events from zero-order Laue zone (ZOLZ) diffraction spots is not feasible. The impact of multiple scattering involving excitation of spots from higher order Laue zones (HOLZs) is negligible in practice, as evidenced by the different visibility of the kinematically forbidden spots in our experimental diffraction patterns along different orientations. Specifically, the (002) spot in the [110] orientation, which can be induced by a combination of elastic scattering events to spots from ZOLZ, shows strong contrast, while the (020) spot in the [001] orientation, which requires a combination from HOLZ, is almost invisible.

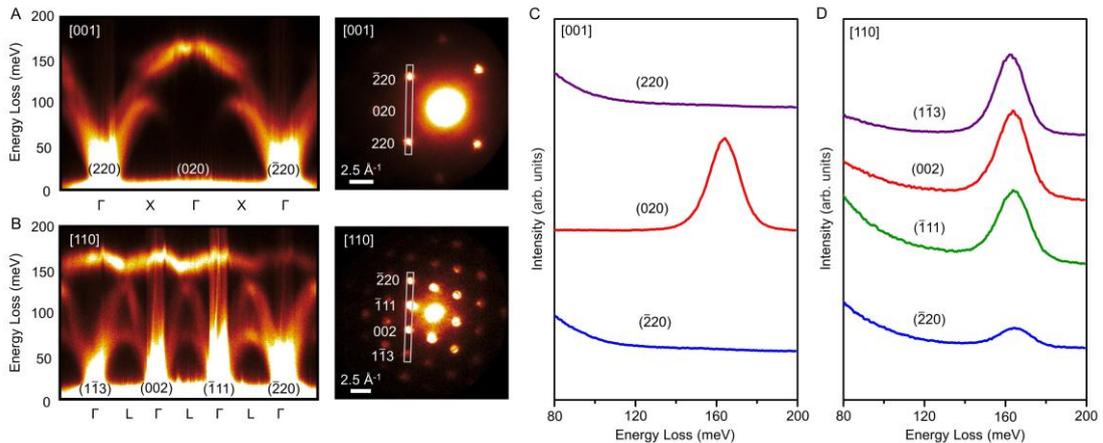

**Figure 3 Momentum-resolved vibrational spectroscopy of diamond.** Phonon dispersions and diffraction patterns obtained along the (A) [001] and (B) [110] axes, respectively. The selected momentum space regions in the diffraction patterns are highlighted. (C) and (D) are the vibrational spectra extracted from the Γ points in (A) and (B), respectively.

To confirm the role of multiple scattering in the visibility of the optical phonon peak at the ($\bar{2}$20) point, we first performed FRFPMS simulations on a 50 nm thick diamond specimen (Fig. 4 and Fig. S5). FRFPMS includes the effects of dynamical diffraction along with the inelastic phonon scattering events. Therefore, if visibility of the optical phonon peak at the ($\bar{2}$20) spot is indeed arising from multiple scattering, FRFPMS simulation should capture this behavior. As shown in the phonon dispersion (Fig. 4A) and the corresponding simulated spectra of the $\Gamma$ points (Fig. 4B) in the [001] orientation, a clear and strong optical phonon peak (around 165 meV) appears at the (020) spot, while it becomes negligible at the (220) and ($\bar{2}$20) spots. These results are highly consistent with our experiment, except that minuscule optical phonon peaks are present in the simulated spectra of the (220) and ($\bar{2}$20) spots. These extremely weak optical phonon peaks at the (220) and ($\bar{2}$20) spots, as well as the weak but nonzero intensity of the simulated zero-loss peak (ZLP) in the spectrum of the kinematically forbidden spot (020), should be attributed to multiple scattering events from HOLZs and are likely below detection limits in our experiments as we mentioned above. In the case of diamond [110] orientation, Figs. 4D-E shows that the optical phonon peak intensities for the spots (1$\bar{1}$3), (002) and ($\bar{1}$11) are very similar, while the optical phonon peak at the ($\bar{2}$20) spot is approximately three times lower but still visible which is in agreement with experiment.

Furthermore, we have also simulated the phonon dispersions for diamond at other thicknesses in both [001] and [110] orientations. Given that the extinction distance of diamond at the (111) spot under 60 kV is approximately 38 nm [46], we should qualitatively expect nonlinear increases in the [110] orientation with varying sample thickness due to multiple scattering effects. This is exactly what our simulations show in Figs. 4C and F: as sample thickness increases from 10 nm to 60 nm, the optical phonon signals at the ($\bar{2}$20) spot remain small and increase linearly in the [001] orientation, while they exhibit significant intensities and nonlinear growth in the [110] orientation, highlighted by the deviations between the fractional intensities at different thickness and their linear fits. For instance, in the latter case, the signal intensity for a 10 nm thick sample is only 1/10$^{th}$ of that for a 50 nm thick sample. Hence, our experimental and simulated results demonstrate that dynamical effects should be considered when interpreting experimental vibrational spectroscopy data, especially for thick samples.

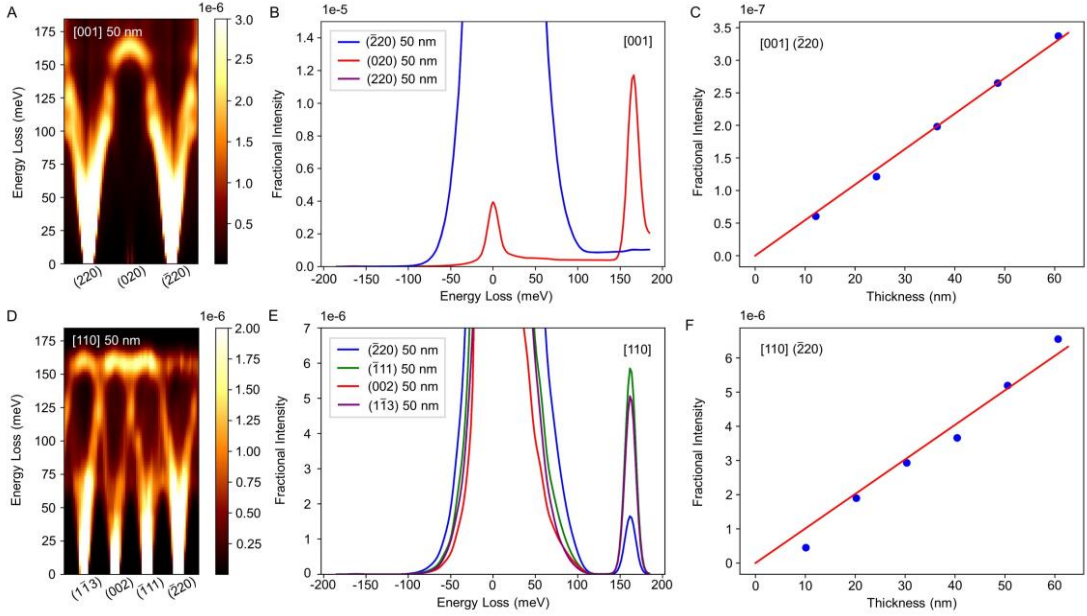

**Figure 4 Simulations of momentum-resolved vibrational spectroscopy of diamond.** Simulated phonon dispersion plots for 50 nm thick samples in the (A) [001] and (D) [110] orientations. (B) and (E) are simulated spectra of Bragg spots corresponding to (A) and (D), respectively. (C) and (F) are the background-subtracted raw (unbroadened) fractional intensities of the optical phonon signals at the ($\bar{2}$20) spot in the [001] and [110] orientations, respectively. The red line in each panel is linear fit of the data.

In summary, our systematic investigation of vibrational signals in graphene, complemented with theoretical analysis, has unveiled a phenomenon of systematic absences in optical phonon modes that originate from destructive interference of the inelastically scattered waves from different basis atoms. Furthermore, our analysis on diamond has confirmed these systematic absences at Γ points as a general characteristic in materials composed of symmetrically equivalent pair(s) of the same elements. The systematic absences also indicate that the vibrational signals of the same phonon modes from different BZs are not the same as those from the first BZ, which is usually neglected in the previous studies. Moreover, it is worth noting that the intensities of vibrational EELS in bulk samples are also modulated by dynamical effects, which is evidenced by the vibrational analysis on the diamond samples with a relatively large thickness. Our study demonstrates the importance of the systematic absences and dynamical effects in the modulation of the vibrational signals. Therefore, it is necessary to consider their impact on the variations of vibrational spectroscopy at the local structures of materials for a comprehensive understanding of their vibrational properties, especially for thick samples.

**Acknowledgement:** This research was supported by the Beijing Outstanding Young Scientist Program (BJJWZYJH01201914430039) and the CAS Project for Young Scientists in Basic Research (YSBR-003). We acknowledge the Swedish Research Council (grant no. 2021-03848), Olle Engkvist's Foundation (grant no. 214-0331), and Knut and Alice Wallenberg Foundation (grant no. 2022.0079) for financial support. The simulations were enabled by resources provided by the National Academic Infrastructure for Supercomputing in Sweden (NAISS) at NSC Centre partially funded by the Swedish Research Council through grant agreement no.2022-06725. This research benefited from resources and supports from the Electron Microscopy Center at the University of Chinese Academy of Sciences. We thank Dr. Tracy Lovejoy for valuable suggestions on momentum-resolved vibrational EELS experiments.

**Author contributions:** W.Z. conceived and supervised the project. A.L. carried out the momentum-resolved vibrational EELS experiments and data analysis with the help of M.X. M.X. and A.L. transferred the graphene sample and M.X. fabricated the diamond lamellas by FIB. P.Z., Z.H. and J.R. performed the FRFPMS simulations and developed the theoretical explanation of systematic absences of optical phonon peaks. A.L., P.Z., J.R. and W.Z. wrote the paper with the input from M.X. and S.J.P.. All authors discussed the results and commented on the manuscript.

# Supplementary Materials for

# Systematic Absences of Optical Phonon Modes in Phonon Dispersion Measured by Electron Microscopy

Aowen Li (李傲雯), Paul Zeiger, Zuxian He (何祖贤), Mingquan Xu (许名权),

Stephen J. Pennycook, Ján Rusz[*], Wu Zhou (周武)[*]

**This PDF file includes:**

Materials and Methods

Additional Figures S1 to S5

Additional References

.

**Materials and Methods**

S1. Sample preparation

The graphene sample used in the momentum-resolved vibrational spectroscopy in STEM was grown by chemical-vapor deposition (CVD) method [1], and then transferred to a $Si_3N_4$ MEMS-based heating chip [2]. The diamond lamellas for cross-sectional investigation in the momentum-resolved vibrational spectroscopy in STEM was prepared using a Thermo Scientific Helios G4 CX DualBeam system. The working voltage and ion beam current for thinning the lamellas was decreased gradually to minimize the sample damage during preparation. The final polishing was performed with the parameters of ion beam of 2 kV and 23 pA to reduce the surface amorphization. Before being loaded into the microscope, the heating chip sample and diamond samples were baked about 18 hours under 160 degrees in vacuum to avoid hydrocarbon contamination. The graphene sample was heated to 550 °C during the momentum-resolved vibrational spectroscopy experiments for a large-scale clean region, as shown in Fig. S1A, while the diamond samples were measured at room temperature.

S2. Momentum-resolved vibrational spectroscopy experiments and data processing

The momentum-resolved vibrational spectroscopy in STEM were conducted under an accelerating voltage of 60 kV using a Nion HERMES-100 microscope equipped with a direct electron detector for EELS [3]. The electron beam was monochromated to 7 pA with a convergence semi-angle of 3.5 mrad. Since the spherical aberration can be ignored at such angles, the probe size $d_r$ was dependent on the diffraction limit $d_d$ and the beam source size $d_g$:

$$d_r = (d_g^2 + d_d^2)^{0.5} = \left\{\left[2(I_p/B)^{0.5}(1/\pi\alpha)\right]^2 + (0.61\lambda/\alpha)^2\right\}^{0.5} \quad (1)$$

where $\alpha$ is the convergence semi-angle, $\lambda$ is the wavelength, $I_p$ is the beam current and $B$ is the source brightness [4]. The diffraction spot radius was calculated by $2\pi\alpha/\lambda$. Therefore, this experimental setup yields a probe size of ~1.2 nm and a diffraction spot radius of 0.45 Å$^{-1}$. Since the vibrational signals of single-layered graphene are extremely weak, especially within the momentum space beyond the first

BZ, we employed a slot-type EELS aperture with a size of 125 μm × 2 μm to perform high-efficiency parallel acquisitions [5]. We also used a state-of-the-art direct electron EELS detector to record graphene phonon dispersions without readout noise [3,6]. Owing to the aforementioned technical advancements, we can feasibly obtain high quality phonon dispersions of monolayer graphene across several BZs. In our experiments, we collected a three-dimensional EELS data consisting of 150 frames of two-dimensional EELS spectra with a dwell time of 5 s for each frame for the phonon dispersions of graphene, and 500 frames with a dwell time of 1 s for those of diamond and h-BN. To obtain vibrational signals from different regions of the momentum space, we manipulated a combination of projector lenses to displace the selected diffraction spots of graphene into the entrance aperture of the spectrometer. For each momentum space region of graphene, we also acquired EELS data at a contaminated region with the same experimental setup to guide the energy alignment of the data (see below). The energy dispersion of the EELS measurements is 0.6 meV per channel.

S3. EELS Data processing

The collected three-dimensional EELS datasets were first aligned and then integrated to a two-dimensional phonon dispersion diagram by the rigid registration method in the Nion Swift software [7]. Since the diffuse scattering is very weak from the monolayer graphene, the zero-loss peaks (ZLP) signals almost vanish between the Γ points, making it difficult to align the energy-loss signals along the non-dispersion direction. The shifts of ZLP in each line of the two-dimensional data are mainly caused by the aberrations in the EELS spectrometer, which should be nearly constant when keeping the experimental setup unchanged. Therefore, we collected EELS data on the contaminated region in the same momentum space and then determined the shifts of the ZLPs line-by-line by Gaussian peak fitting in the two-dimensional data. Finally, we corrected the corresponding phonon dispersion diagram of graphene by the measured values of shifts. For the data collected from the diamond samples, the phonon dispersions were aligned by the ZLPs of the data itself. The correction of the shifts of the ZLPs in a two-dimensional phonon dispersion diagram was performed using custom-written python code. In the corrected data, the energy resolution (the FWHMs of ZLPs) of both the spectra at the middle Γ point in Fig. 1A and the ($\bar{1}11$) spot in Fig. 3B is 16 meV and become slightly worse when the momentum transfer of incident electron is increased. The thicknesses of the diamond samples were calculated by the log-ratio method in the DigitalMicrograph software [8].

S4. Theoretical derivation of the systematic absences of optical phonon peaks in graphene

For incident electrons scattered to a specific Γ point ($hk0$) of graphene, their momentum transfer can be represented as:

$$\boldsymbol{q} = \boldsymbol{0} + h\boldsymbol{b}_1 + k\boldsymbol{b}_2 \quad (2)$$

where $\boldsymbol{b}_1$ and $\boldsymbol{b}_2$ are reciprocal lattice vectors and $\boldsymbol{0}$ represents the central Γ point. Additionally, the positions of the two basis atoms of graphene can be described as follows:

$$\boldsymbol{R}_1 = 0\boldsymbol{a}_1 + 0\boldsymbol{a}_2 \quad (3a)$$

$$\boldsymbol{R}_2 = \frac{1}{3}\boldsymbol{a}_1 + \frac{2}{3}\boldsymbol{a}_2 \quad (3b)$$

where $\boldsymbol{a}_i$ ($i$=1, 2) are the lattice vectors that satisfy $\boldsymbol{a}_i \cdot \boldsymbol{b}_j = 2\pi\delta_{ij}$. Therefore, the second factor of Eq. (3) in the Main Text can be rewritten as follows:

$$\left|e^{-i(h\boldsymbol{b}_1+k\boldsymbol{b}_2)\cdot\boldsymbol{R}_1} - e^{-i(h\boldsymbol{b}_1+k\boldsymbol{b}_2)\cdot\boldsymbol{R}_2}\right|^2 = 2\left[1 - \cos\left(\frac{2\pi(h+2k)}{3}\right)\right] \quad (4)$$

S5. Theoretical derivation of the systematic absences of optical phonon peaks in diamond

The double differential scattering cross-section of the diamond optical phonons at a Γ point in diamond ($\boldsymbol{q} = h\boldsymbol{b}_1 + k\boldsymbol{b}_2 + l\boldsymbol{b}_3$, where $\boldsymbol{b}_1, \boldsymbol{b}_2,$ and $\boldsymbol{b}_3$ are reciprocal lattice vectors defined by the conventional cell) is proportional to:

$$\frac{d^2\sigma}{d\Omega(\boldsymbol{q})dE} \propto \left|\left[1 + e^{-i\pi(h+k)} + e^{-i\pi(k+l)} + e^{-i\pi(h+l)}\right] \cdot \left[1 - e^{-i2\pi(h+k+l)/4}\right]\right|^2 \quad (5)$$

Therefore, the condition for Eq. (5) to vanish is $h + k + l = 4n$ ($h$, $k$, $l$, and $n$ are integers) or that $h$, $k$, and $l$ are mixed odd and even integers. This condition is similar to the extinction rule for diamond in electron diffraction (*i.e.* $h + k + l = 4n + 2$ or $h$, $k$, and $l$ are mixed odd and even integers). Because under the conventional cell notation, all Γ points satisfy the condition that $h$, $k$, and $l$ are all odd or even, the rule of the systematic absences of optical phonon peaks in diamond is $h + k + l = 4n$.

S6. FRFPMS simulation

Molecular dynamics simulations of graphene were performed with LAMMPS [9] using a GAP machine-learning interatomic potential [10]. Simulation cell dimensions were 4.696 nm × 4.683 nm × 1.0 nm, containing 832 carbon atoms forming a single sheet of graphene surrounded by 0.5 nm of vacuum on both sides. Periodic boundary conditions were used. Temperature was set to 550 °C and a canonical ensemble was simulated using a Langevin thermostat with a timestep of 1 fs and damping constant of 0.1 ps. After 5 ps of thermalization, 200 ps of trajectory data was accumulated, which was subsequently used to generate snapshots of vibrating structure. The resulting trajectory data was repeatedly band-pass filtered, so that atomic motion only within a desired range of vibrational frequencies remained in the filtered trajectory. In this way snapshots for a FRFPMS simulations have been generated. We have used a grid of frequencies from 2.0 THz up to 50.0 THz with a step of 1 THz. In each frequency bin 100 snapshots have been generated and they were used as an input for multi-slice calculations using DrProbe [11]. We have used a $B_{iso}$ parameter of 0.00276 nm$^2$ to account for Debye-Waller smearing [12]. A real-space numerical grid of 836 × 836 pixels was used. The electron beam acceleration voltage is 60 kV. Both the convergence semi-angle and collection semi-angle applied in the simulations are 3.5 mrad, following experimental settings. Single slice along the z-direction has been considered. The simulated spectra shown in Fig. 2B are normalized to the bin at around 100 meV. All spectra are convolved with a Gaussian of full widths at half maximum (FWHM) of 16 meV, matching the width of the experimental ZLP.

For diamond we have used a mostly analogous procedure for both orientations. First the diamond lattice constant was determined to be 3.57329 Å at zero pressure and at a temperature of 300 K for the Tersoff interatomic potential [13,14]. We have set up two structure models, one for each of the considered [001] and [110] orientations. The size of the simulation box was 10 × 10 × 170 conventional unit cells (3.57329 nm × 3.57329 nm × 60.74594 nm) for the [001] orientation and a size of 7 × 10 × 240 unit cells (3.53738 nm × 3.57329 nm × 60.64082 nm) for the [110] orientation. The time step was set to 0.5 fs in all MD simulations for diamond. We ran a constant temperature MD simulation using a Langevin thermostat and sampled the velocities $v$ and positions $R$ of all atoms every 10,000 time steps after an initial equilibration of 20000 time steps for a total of 40 such samples. These samples were used as the initial conditions for 40 trajectories of constant energy MD simulations of a length of 100,000 time steps each after discarding the first 10000 steps of each trajectory. These trajectories were subsequently repeatedly band-pass filtered in order to sample snapshots for a FRFPMS simulation. For both orientations, the grid of frequencies was set to a range from 0.5 THz up to 45.0 THz with a step of 0.5 THz, in total 90 frequency bins. In total 200 snapshots have been calculated in each frequency bin. In the [001] orientation, the real-space grid was set to 560 × 560 pixels and total number of slices along the z-direction was 1360. In the [110] orientation, the real-space grid was set to the same size 560 × 560 pixels as in the [001] orientation, however, the number of slices was 960. Five different sample thicknesses have been calculated with a step of ~12 nm up to ~60 nm. For both orientations, the number of slices was chosen to be an integer multiple of number of unit cells along the z-axis times the number of planes containing carbon atoms in respective orientations. For this reason, different numbers of slices were chosen in each of the two orientations. In all simulations the Debye-Waller factor was set using $B_{iso}$ parameter of 0.00087 nm$^2$. An acceleration voltage was set to 60 kV and convergence semi-angle to 3.5 mrad, following experimental settings.

Once FRFPMS datasets of inelastic scattering intensity as a function of momentum transfer $\mathbf{q}$ and energy loss $E$ were obtained, spectra were assembled using Eq. 48 of Ref. [12], including also a ZLP based on the calculated coherent (elastic) intensity. These spectra were subsequently broadened by a convolution with a Gaussian function with FWHM corresponding to experimental ZLP FWHM in vacuum. The simulated spectra for the (220) and ($\bar{2}$20) spots in the [001] orientation are visually indistinguishable at this scale, suggesting a high level of convergence of the simulation.

The fractional intensities of the optical phonon signals at the ($\bar{2}$20) spot with different thicknesses shown in Figs. 4C and F are obtained by using statistics-sensitive non-linear iterative peak-clipping (SNIP) algorithm for background subtraction in the optical phonon peak region of the raw (unbroadened) simulated spectra [15].

**Additional Figures:**

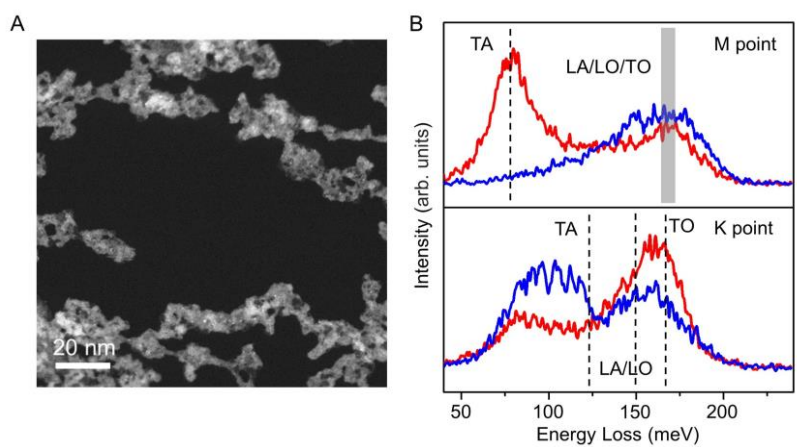

**Figure S1. Momentum-resolved vibrational spectroscopy of graphene at the M and K points.** (A) High-angle annular dark-field (HAADF) image of graphene. (B) Vibrational spectra extracted from the highlighted M and K points in Figs. 1A-B. The dashed lines and gray rectangle in this figure indicate the energies of the phonon modes at the corresponding high-symmetry points.

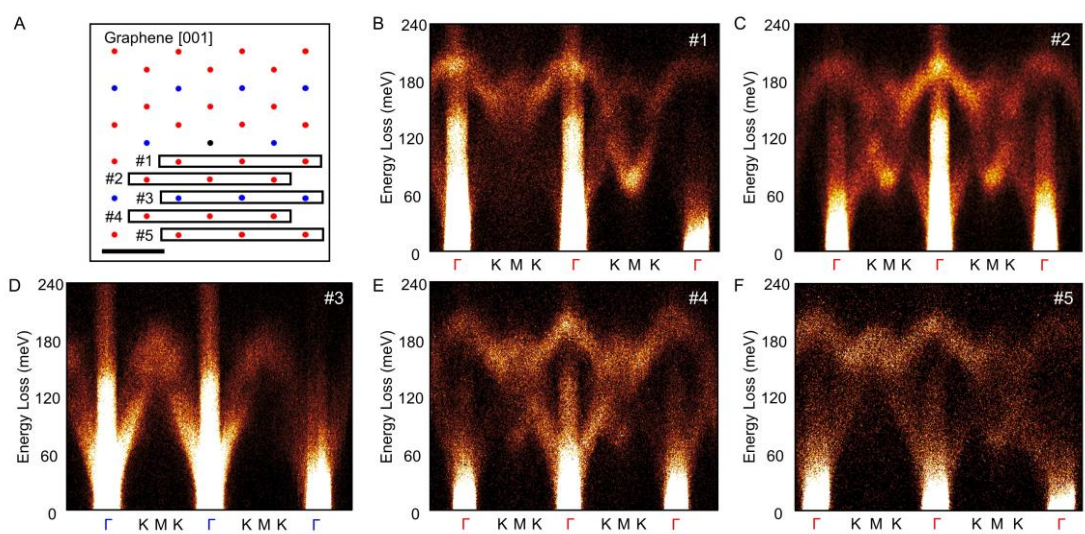

**Figure S2. Phonon dispersion of graphene obtained at different momentum space regions.** (A) Schematic of graphene diffraction pattern along the [001] zone axis. Scale bar: 5 Å$^{-1}$. The selected momentum space regions #1-#5 are denoted. (B-F) The phonon dispersion diagrams obtained from the corresponding momentum regions in (A).

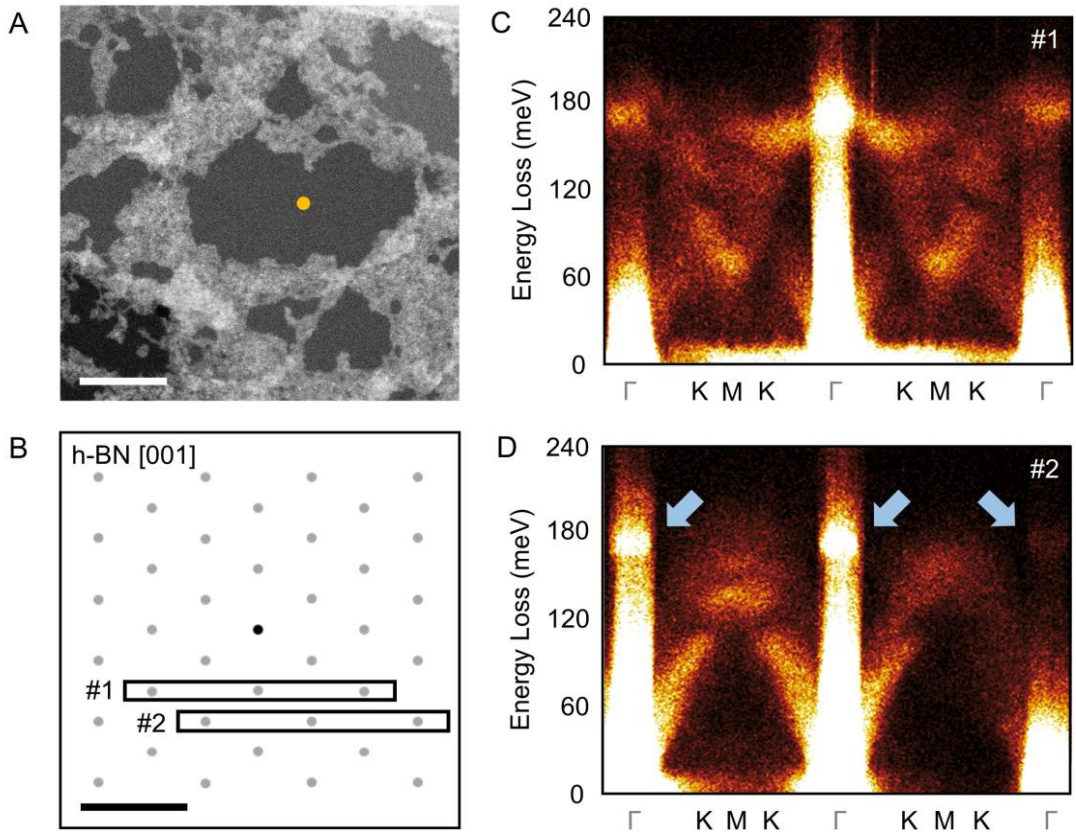

**Figure S3. Phonon dispersion of h-BN obtained at different momentum space regions.** (A) Medium-angle annular dark-field (MAADF) image of h-BN. An orange dot indicates the region for collecting phonon dispersion diagrams, which is four-layer thick. Scale bar: 20 nm. (B) Schematic of h-BN diffraction pattern along the [001] zone axis. Scale bar: 5 Å$^{-1}$. The selected momentum space regions #1 and #2 are denoted. (C-D) The phonon dispersion diagrams obtained from the corresponding momentum regions in (B). The ZLPs in (C) were aligned based on the data itself, while those in (D) were aligned by the corresponding data on contamination.

The indices of the Γ points in Fig. S3D satisfy $h + 2k = 3n$ ($n$ denoted integers). Nevertheless, there are strong signals appearing at these Γ points, highlighted by the blue arrows, which should be attributed to the surface phonon polaritons (SPhPs) of h-BN.

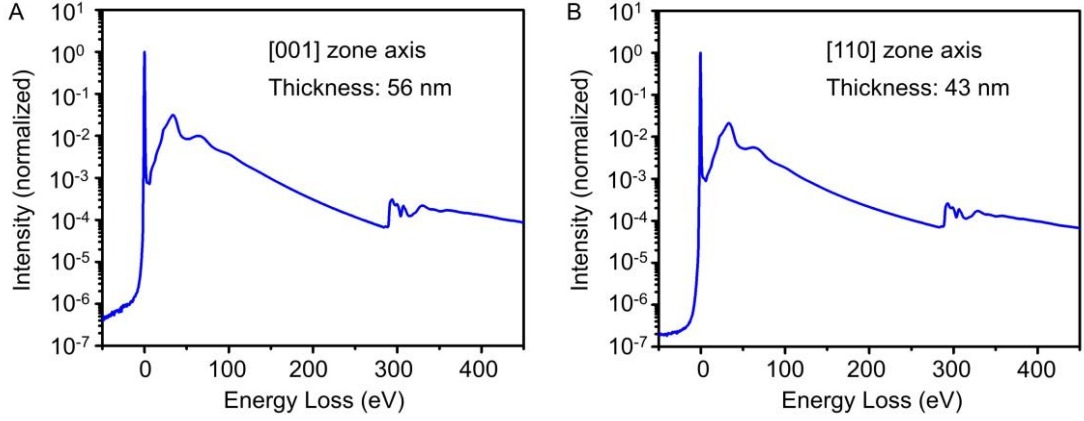

**Figure S4. Thickness measurements of the diamond samples.** The thickness of the diamond samples along the [001] and [110] zone axes are (A) 56 and (B) 43 nm, respectively. Both spectra are normalized by the maximums of their ZLPs.

The magnitude of extinction distance ($\xi_g$) of diamond can be expressed as follows [16]:

$$\xi_g \propto \frac{\pi V_c \cos\theta_B}{\lambda |F_g|} \qquad (6)$$

where $V_c$ is the volume of unit cell, $\theta_B$ is the Bragg angle, $\lambda$ is the wavelength of fast electron and $F_g$ is the structure factor. For instance, according to this formula, the extinction distance of the (111) spot of diamond at 60 kV is about 38 nm. Therefore, dynamical effects are not negligible when considering the interaction between the fast electrons and our 50 nm-thick diamond samples, explaining the nonlinear relationship between the optical phonon peak intensity and the sample thickness in our theoretical simulations.

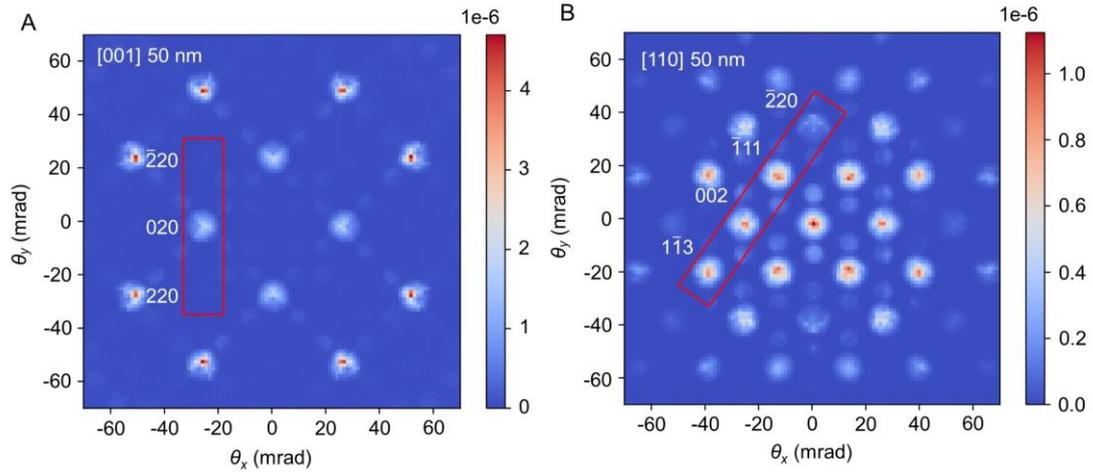

**Figure S5. Simulated energy filtered diffraction patterns of the diamond samples.** The energy filtered diffraction patterns showing excitations of phonons around 165 meV for (A) [001] and (B) [110] orientation. This frequency corresponds to the position of the optical phonon peak. Red rectangle mark areas from which the simulated phonon dispersion in Figs. 4A and D has been extracted, following the area measured in experiment as marked in Figs. 3A and B. The thicknesses of the diamond samples in the simulations are 50 nm.